\documentclass[%
 reprint,twocolumn,
 notitlepage,showkeys,
superscriptaddress,nofootinbib,
 amsmath,amssymb,
 aps,
prd,
]{revtex4-2}
\usepackage{epsfig}
\usepackage{hyperref}
\usepackage{orcidlink}
\hypersetup{           
    colorlinks=true,                
    breaklinks=true,                
    urlcolor= cyan,                
    linkcolor= red,                
    bookmarksopen=false,
    filecolor=black,
    citecolor=blue,
    linkbordercolor=blue}
\usepackage{graphicx}
\usepackage{orcidlink}
\usepackage{multirow}
\usepackage[utf8]{inputenc}

\begin{document}
\title{Emergent Cosmology in 4D Einstein Gauss Bonnet Theory of Gravity}

\author{Mrinnoy M. Gohain\orcidlink{0000-0002-1097-2124}}
\email{mrinmoygohain19@gmail.com}
\thanks{(Corresponding Author)}
\affiliation{%
 Department of Physics, Dibrugarh University, Dibrugarh \\
 Assam, India, 786004}

\author{Kalyan Bhuyan\orcidlink{0000-0002-8896-7691}}%
 \email{kalyanbhuyan@dibru.ac.in}
\affiliation{%
 Department of Physics, Dibrugarh University, Dibrugarh \\
 Assam, India, 786004}%

\keywords{4D-Einstein-Gauss-Bonnet Gravity, Cosmology, Emergent Cosmology, Modified Gravity, Einstein Static Universe}
\begin{abstract}
In this paper, in an FLRW background and a perfect fluid equation of state, we explore the possibility of the realization of an emergent scenario in a 4D regularized extension of Einstein-Gauss-Bonnet gravity, with the field equations particularly expressed in terms of scalar-tensor degrees of freedom. By assuming non-zero spatial curvature ($k = \pm 1$), the stability of the Einstein static universe (ESU) and its subsequent exit into the standard inflationary system is tested through different approaches. In terms of dynamical systems, a spatially closed universe rather than an open universe shows appealing behaviour to exhibit a graceful transition from the ESU to standard cosmological history. We found that under linear homogeneous perturbations, for some constraints imposed on the model parameters, the ESU is stable under those perturbations. Moreover, it is noted that for a successful graceful transition, the equation of state $\omega$ must satisfy the conditions $-1 < \omega <0$ and $\omega < -1$ for closed and open universes, respectively. Furthermore, the ESU is seen to be neutrally stable under matter perturbation in the Newtonian gauge.
\end{abstract}

\maketitle

\section{Introduction}
\label{intro}
The theory of general relativity suggests that at the genesis of the Universe, spacetime and matter were compressed to a region of infinitesimally pointlike singularity (the initial singularity) with infinite density, which in other words is also known as the Planck's scale. At Planck's scale, the known classical laws of physics break down and quantum mechanics becomes significantly important. In an attempt to unravel the nature of the initial singularity, the incompetence of classical general relativity led to the development of alternative theories, or to be precise theories of quantum gravity which are based on the application of quantum mechanics in gravity. Thus quantum gravity (and hence quantum cosmology) has proven to be quite popular in resolving the problem of the initial singularity. Apart from this, string theory, ekpyrotic/cyclic, and bouncing universe theories are also some of the candidates proposed to address the initial singularity problem. In addition, a relatively new idea to settling the debate of the problem of initial singularity, Ellis et al. \cite{ellis1, ellis2} suggested the so-called ``\emph{emergent Universe}" (EU) scenario in the framework of general relativity and closed Friedmann-Lemaitre-Robertson-Walker (FLRW) universe. The EU is, therefore, a singularity-free Universe where the Universe is supposed to be ever-existing in an Einstein static (ES) phase and subsequently undergo a phase transition into the standard inflationary domain. Therefore, in an EU, the initial singularity is replaced by the ESU. The original Ellis et al. model \cite{ellis1,ellis2}, which was developed in general relativity, faced significant fine-tuning problems. The authors suggested that, rather than a Big Bang singularity, an initially static state known as the ESU existed in the eternal past, following a closed FLRW cosmological setting with positive spatial curvature, the exit from which leads to a brief phase of the inflationary era and then undergo reheating the usual way. In their original model, they formulated a possible non-singular contender of a conventional inflationary Universe with a singularity by considering the feasibility of an early Universe with positive spatial curvature. In particular, they modelled the scenario with a physically viable potential and a scalar field. The presence of the initial static Universe solves the so-called ``horizon problem" as it turns out, this configuration naturally obviates the horizon problem.  Consequently, a self-consistent, or stable, departure from the ESU and a graceful transition from the ESU to the inflationary phase are necessary for the scenario to successfully describe the fixing of the initial singularity. Notably, the former and latter are sufficient and required criteria for singularity evasion, indicating that the EU scenario fails if any of the two requirements are not satisfied. Ellis et al's model failed to achieve the former criteria, posing a significant obstacle to stability. The original EU scenario failed to successfully resolve the big bang singularity issue given that Barrow et al \cite{Barrow2003May} discovered that the ESU in GR is not stable, suggesting that the universe in such an initial static state cannot survive for long against perturbations in the ES phase. However, in the early universe, physical situations in particular, like gravity quantization or GR-based corrections may tip the balance in favour of the EU scenario. In a nutshell, although the EU scenario collapsed in the context of GR, current modified gravity theories could potentially able to ameliorate the situation. This idea has prompted various investigations on the natural improvements of the original EU setup into modified gravity theories, intending to achieve a few promising findings in contrast to GR. 
In the aforementioned context, it seems that the issues of stability are resolved while working on modified theories of gravity\cite{parisi}. With this motive, the ESU is extensively studied in \cite{Bohmer, seahra,shabani}, which led to interesting properties that are substantially dissimilar to those of general relativity, when stability is concerned. In agreement with this statement, we take into consideration the possibility of a modified matter-geometry scenario in the case of an EU scenario with a motive to account for the stability of the ESU.  Therefore, it is quite crucial to address the stability of a model to avoid the collapse of the ESU into the initial singularity. The stability analysis for a stable ESU can be done in terms of different methods like homogeneous, inhomogeneous perturbations and also in terms of dynamical systems. Some of these methods are well described in Ref. \cite{Huang2015Jan,Khodadi2022Jun,Atazadeh2017Jun,Heydarzade2015Apr,Sharif2019Dec,
Mousavi2017Jun,Hadi2018Jan,Darabi2018Jul,Huang2020Sep,Shabani2022Sep,
Darabi2015Aug,Bohmer2015Dec}. Emergent cosmology is an approach to address the problem of the big-bang singularity, with a modification of the standard inflationary Universe, with an ever-existing phase of ESU, with a radius greater than Planck's scale to avoid the quantum gravity era. The ESU has been revisited as a potential foundation for an EU, which is dedicated to addressing the avoidance of the initial singularity prevalent within inflationary cosmology. The ESU exhibit notable features, including the absence of an initial singularity and mitigation of the issues associated with quantum gravity. It is important to emphasize that the stability of the ESU plays a pivotal role in the successful implementation of the EU scenario. Extensive studies on the stability of the ESU have been carried out within various modified gravity theories as well as in theories with various physically motivated corrections in different theories. These studies can be found in Ref. \cite{Parisi2012Jul,Ghorani2021Jun,Li2017Jul,Zhang2016Jul,Huang2015May,
Khodadi2017Dec,Guendelman2015Nov,Miao2016Oct,Khodadi2015Dec,
Huang2018Jan,Shabani2019Mar,Li2019May,Heydarzade2016Jun,Khodadi2016Jun,
Khodadi2018Jun,Paul2018Feb,Gangopadhyay2016Oct,Huang2015Jan,Khodadi2022Jun,
Atazadeh2017Jun,Heydarzade2015Apr,Sharif2019Dec,Mousavi2017Jun,Hadi2018Jan,
Darabi2018Jul,Huang2020Sep,Guendelman2024Jan,Shabani2022Sep,Paul2020Aug,
Darabi2015Aug,Bohmer2015Dec,Gohain2023Mar,Gohain2024Feb}

With these motivations at hand, the paper is dedicated to the possibility and consequences of emergent cosmology in the framework of 4D-Einstein-Gauss-Bonnet gravity, which is planned as follows: In section \ref{sec2} we discuss and review the basic formulation of $4D$-Einstein Gauss-Bonnet gravity. In section \ref{sec3} we study the stability and graceful exit mechanism of emergent cosmology using dynamical systems. In section\ref{sec4}, we study the stability of the ESU subject to homogeneous linear perturbations. In section \ref{sec5}, we discuss the consequences of matter perturbations in the Newtonian gauge to the ESU and finally in section \ref{conc} we summarize and conclude the outcomes of the study.

\section{Brief Review of 4DEGB Gravity}
\label{sec2}
Fernandes et al.\cite{Fernandes2020Jul} have proposed an innovative regularisation method for Einstein Gauss-Bonnet gravity, resulting in a set of field equations that can be stated in closed form in 4D. Their approach demands the inclusion of a counter-term to the action term and is unaffected by the embedding and compactification of any higher-dimensional space. This counter-term eliminates the divergence in the action that would otherwise occur. For detailed derivation see Ref. \cite{Fernandes2020Jul,Fernandes2021May,Fernandes2022Feb,LuPang,
Hennigar2020Jul}
.Moreover, from an observational perspective Clifton et al \cite{Clifton2020Jun}studied the observational constraints on the 4DEGB gravity and Toniato et al \cite{Toniato2024Feb} developed a complete post-Newtonian analysis of 4DEGB theories, improving some observational constraints and also discussed extensions to the PPN formalism accordingly.
 The action resulting from the addition of the Gauss-Bonnet term in the Einstein-Hilbert action in $D$ dimensions with the Glavan-Lin rescaling $\alpha \to \frac{\alpha}{D-4}$ is \cite{Fernandes2020Jul}
\begin{equation}
S = \int d^D x \sqrt{-g} \left(R + \frac{\alpha}{D-4} \mathcal{G}\right)
\label{act_D}
\end{equation}
where $\mathcal{G}$ is the Gauss-Bonnet scalar given by 
\begin{equation}
\mathcal{G} = R^2 - 4R_{\mu \nu}R^{\mu \nu} + R_{\mu\nu\rho\sigma}R^{\mu\nu\rho\sigma},
\label{GBS}
\end{equation}
It may be noted that the rescaling of the Gauss-Bonnet coupling constant, is a way of accounting for the so-called \emph{conformal} or \emph{trace} anomaly in quantum field theory \cite{Fernandes2022Feb}.
Following \cite{Fernandes2020Jul} directly, one may write the resulting action after $4D$-regularization first reported by Lu and Pang \cite{LuPang}
\begin{equation}
\begin{aligned}
& S=\int_{\mathcal{M}} d^D x \sqrt{-g}\left[R+ \alpha(D-4)\left(4(D-3) G^{\mu \nu} \nabla_\mu \phi \nabla_\nu \phi \right.\right. \\
&\left.\left. -\phi \mathcal{G}-4(D-5)(D-3) \square \phi(\nabla \phi)^2-(D-5)(D-3)\right.\right. \\
&\left.\left. \hspace{4.5cm}(D-2)(\nabla \phi)^4\right)\right]+S_m
\end{aligned}
\label{action2}
\end{equation}
where $\phi$ is a scalar degree of freedom associated through a conformal transformation of the metric tensor given by $\tilde{g}_{\mu \nu} = e^{2\phi} g_{\mu \nu}$.
In the $4-D$ limit equation \eqref{action2} reduces to 
\begin{equation}
\begin{aligned}
S&=\int_{\mathcal{M}} \mathrm{d}^4 x \sqrt{-g}\left[R+\alpha\left(4 G^{\mu \nu} \nabla_\mu \phi \nabla_\nu \phi-\phi \mathcal{G} \right.\right. \\
&\left.\left. \hspace{3cm}+4 \square \phi(\nabla \phi)^2+2(\nabla \phi)^4\right)\right]+S_M,
\end{aligned}
\label{action3}
\end{equation}
which is a $4-D$ action free of divergences. 

The 4D regularized Einstein Gauss Bonnet theory has several interesting implications for instance, in higher dimensions, it agrees rather well with the Gauss-Bonnet theory's Kaluza-Klein reduction \cite{LuPang,Kobayashi2020Jul}. In addition, it also appears to be exactly compatible with the effective action that results from the trace anomaly in quantum theory, which is caused by the broken conformal symmetry of massless fields \cite{Kobayashi2020Jul,Komargodski2011Dec}. This may imply that the theory may be interpreted as a gravitational theory with an established quantum correction mechanism plus a theory with reduced dimensionality. From a phenomenological perspective, the theory appears to be in the absence of Ostrogradski instabilities and also permits intriguing cosmological and black hole solutions that essentially reduce to the equations in GR. The accelerating behaviour that the cosmic solutions offer may be of significance in the high- and low-redshift domains \cite{Fernandes2022Feb} (and references therein).

Let us now consider the homogeneous and isotropic FLRW metric in $D$ dimensions
\begin{equation}
ds^2 = - dt^2 + a^2 (t) \left[d\chi^2 + S_k^2 (\chi) d\Omega^2 \right],
\label{flrw}
\end{equation}
where $a(t)$ is the scale-factor, 
$$ S_k (\chi) = \left \{\begin{matrix}
\chi, \hspace{1.8cm} k = 0\\
\sin (\chi), \hspace{1.2cm} k = 1 \\
\hspace{0.3cm}\sinh (\chi), \hspace{1cm} k = -1
\end{matrix}
\right.
$$
and $d\Omega^2$ represents the line-element for a $D-2$ sphere. We assume a perfect fluid energy-momentum tensor $T^{\mu}_{\nu} = \text{diag}(-\rho, p, p, p)$ and the scalar-tensor version of 4DEGB gravity \cite{Fernandes2021May,Fernandes2022Feb} for this particular work. 


The Friedmann equation in the 4DEGB gravity has been obtained in \cite{Fernandes2022Feb}, which is given by
\begin{equation}
H^2+\frac{k}{a^2}+\alpha\left(H^2+\frac{k}{a^2}\right)^2=\frac{8 \pi G}{3} \rho+\frac{\alpha C^2}{a^4}.
\label{freq3}
\end{equation}
which includes a dark-radiation term parameterized by $C$. This is referred to as a dark-radiation term as it evolves like radiation in standard cosmology ($\propto \frac{1}{a^4}$).
This term has important cosmological aspects, for instance, very recently, Zanoletti et al \cite{Zanoletti2024Jan} obtained detailed cosmological aspects of the 4DEGB theory that places empirical constraints based on CMB data on the dark radiation parameter $C$, and also computed the perturbed equations of motion for all values of the curvature parameter $k$. 
It will be interesting to explore, the feasibility of non-singular ES behaviour of the early Universe, which is a crucial element of emergent cosmology. Therefore, the Friedmann equation \eqref{freq3} shall be our primary interest from which the dynamics of emergent cosmology shall be studied. The other counterpart to this equation is the Raychaudhuri equation, which can be obtained by differentiating equation \eqref{freq3} and using the continuity equation $\dot{\rho} + 3 H (\rho + p) = 0$, which gives
\begin{equation}
\begin{aligned}
\left( \dot{H} - \frac{k}{a^2}\right) \left[1 + 2\alpha \left(H^2 + \frac{k}{a^2}\right)\right] &+ \frac{2\alpha C^2}{a^4} \\&= -4\pi G (\rho + p),
\end{aligned}
\label{RCEqn}
\end{equation}
The crucial point to be noted here is, in the case of $k = 0$ (flat Universe), from the Friedmann equation \eqref{freq3} one may see that $\rho_s = - \frac{\alpha C^2}{a_s^4}$, ($\rho_s$ and $a_s$ are the energy density and ES radius respectively), is negative for $\alpha > 0$, which is unphysical. However, $\alpha < 0$ leads to positive energy density, but $\alpha < 0$ is inconvenient for consistent cosmology in \cite{Fernandes2022Feb}. Therefore, in this work, the cases of closed and open Universes ($k = 1$ and $k = -1$) will be investigated. 

The stability analysis of the ESU has been extensively used in Gauss-Bonnet gravity by different authors. For instance, Huang et al performed a stability analysis of the ESU in modified Gauss-Bonnet gravity under scalar perturbations to the Newtonian Gauge through harmonic decomposition of the potentials associated with it \cite{Huang2015Jan}. They found that a closed Universe admits stable ES solutions subject to homogeneous perturbation but are unstable in terms of inhomogeneous perturbations, whereas an open Universe is unstable subject either to homogeneous or inhomogeneous perturbations. More recently, Li et al \cite{Li2020Apr} found stable ES solutions to scalar perturbations in a 4D Gauss-Bonnet gravity by rescaling the Gauss-Bonnet coupling constant and performing the analysis in the $D\to 4$ limit. B\"{o}hmer and Lobo \cite{Bohmer2009Mar} studied the stability of the ESU in the context of linear homogeneous perturbations within modified Gauss-Bonnet gravity, by assuming a general form of the Gauss-Bonnet function characterized by a linear equation of state and the second derivative of the Gauss-Bonnet term. To the best of our knowledge, we find that the graceful exit dynamics, which is a requirement of emergent cosmology has not been explored in the aforementioned works or any other literature, as far as 4DEGB is concerned. Therefore in our work, we consider a 4D-regularized scalar-tensor version of the Gauss-Bonnet field equations and perform the stability analysis in various contexts like dynamical stability, homogeneous scalar perturbation to scale factor and energy density, inhomogeneous density perturbation, vector and tensor perturbations. As stated before, given the requirement of a successful emergent cosmology, we shall also try to address the possibility of graceful exit through the method of dynamical systems and scalar homogeneous perturbations.
\section{Graceful Exit Mechanism and Stability Analysis}
\label{sec3}
Emergent cosmology is based on the assumption that the initial singularity is replaced by a stable ESU. This follows from the basic criteria to be fulfilled for a successful emergent cosmology: \emph{existence of stable and sustained ESU} and \emph{a graceful exit from the stable ESU to the standard cosmology.} 
 In this section, we shall address both the mechanism of phase transition and stability based on dynamical system analysis. 
The stability of a dynamical system is performed based on the linearised system $\dot{x}_i = J_{ij} (x_j - x_{j0})$ around the equilibrium point or critical point $(x_{10}, x_{20}) = (a_s, 0)$, where $a_s$ is the ES radius in our context. $J_{ij}$ represents the elements of the Jacobian $J$ defined as 
\begin{equation}
\begin{aligned}
J = \left(\frac{\partial \mathcal{X}_i}{\partial x_j} \right)_{(a_{ES},0)} &= \begin{pmatrix}
\frac{\partial \mathcal{X}_1}{\partial x_1}\left.\right|_{(a_{ES},0)} & \frac{\partial \mathcal{X}_1}{\partial x_2}\left.\right|_{(a_{ES},0)} \\
\frac{\partial \mathcal{X}_2}{\partial x_1}\left.\right|_{(a_{ES},0)} & \frac{\partial \mathcal{X}_2}{\partial x_2}\left.\right|_{(a_{ES},0)}
\end{pmatrix} \\ &= \begin{pmatrix}
\hspace{-1cm} 0 & 1 \\
\frac{\partial \mathcal{X}_2}{\partial x_1}\left.\right|_{(a_{ES},0)} & 0
\end{pmatrix}.
\end{aligned}
\label{Jmat}
\end{equation}
Using Lyapunov's method, the stability of the critical point $(x_1, x_2) = (a_{ES}, 0)$ is determined by the eigenvalues $\lambda$ of the $J$-matrix (\ref{Jmat}).
The eigen-values of the Jacobian can be obtained by calculating the roots of the characteristic equation
\begin{equation}
\lambda^2 - \lambda \, \mathrm{Tr} ( J ) + \mathrm{Det} (J) = 0,
\label{charac_eqn}
\end{equation}
The roots of the characteristic equation are
\begin{equation}
\lambda_{1,2} = \frac{1}{2} \left[ \mathrm{Tr} (J) \pm \sqrt{(\mathrm{Tr}(J))^2 - 4 \mathrm{Det} (J)} \right].
\label{eig_sol_gen}
\end{equation}
The stability of the critical point can be inferred from the sign of $\lambda^2$, i.e. for $\lambda^2 < 0$ and $\lambda^2 > 0$ the critical points are stable and unstable respectively.

We may now construct the dynamical system for the Raychaudhuri equation \eqref{RCEqn}.
Let us consider $x_1 = a$, $x_2 = \dot{x_1} = \dot{a}$. Thus the dynamical system of equations becomes

\begin{widetext}
\begin{equation}
\begin{aligned}
\dot{x_1} &= x_2 \equiv \mathcal{X}_1 (x_1, x_2), \\
\dot{x_2} &= \frac{-\frac{2 \alpha  C^2}{x_1^3}-4 \pi  G \rho  x_1 (\omega +1)+\frac{2 \alpha  k^2}{x_1^3}+\frac{2 \alpha  k x_2^2}{x_1^2}+\frac{2 \alpha  k x_2^2}{x_1^3}+\frac{k}{x_1}+\frac{2 \alpha  x_2^4}{x_1^4}+x_2}{\frac{2 \alpha  k}{x_1^2}+\frac{2 \alpha  x_2^2}{x_1^2}+1} \equiv \mathcal{X}_2 (x_1, x_2)
\end{aligned}
\end{equation}
\end{widetext}
where we have used the perfect fluid EoS parameter $\omega = \frac{p}{\rho}$.
\subsection{Model 1: \(k = 1\)}
Let us first consider the case of a closed Universe. By
setting $k=1$ in the Raychaudhuri equation \eqref{RCEqn} and considering $x_1 = a $ and $x_2 = \dot{x}_1 = \dot{a}$, we can construct the dynamical system in this model as
\begin{widetext}
\begin{equation}
\begin{aligned}
\dot{x}_1 &= x_2 \equiv \mathcal{X}_1(x_1, x_2),\\
\dot{x}_2 &= \frac{2 \alpha  x_1 \left(-C^2+x_2^2+1\right)-4 \pi  G \rho  x_1^5 (\omega +1)+2 \alpha  x_2^2 x_1^2+2 \alpha  x_2^4+x_2 x_1^4+x_1^3}{2 \alpha  \left(x_2^2+1\right) x_1^2+x_1^4} \equiv \mathcal{X}_2(x_1,x_2).
\end{aligned}
\label{dynsysk=1}
\end{equation}
\end{widetext}
The eigenvalue $\lambda$ for this system about the critical point $(x_1,x_2) = (a_s, 0)$ is 
\begin{equation}
\begin{aligned}
 \lambda^2 &= \frac{1}{2 \left(2 \alpha  a_s+a_s^3\right)^2} \left[4 \alpha  \left(3 C^2-2\right) a_s^2- \right. \\& \left.\left((\omega +1) a_s^4 \rho _s \left(a_s^2+6 \alpha \right)\right)-2 a_s^4+8 \alpha ^2 \left(C^2-1\right) \right]
 \end{aligned}
 \label{lamsqk=1}
\end{equation}

Since the ESR $a_s$ is arbitrary and positive, for the sake of mathematical simplicity, let us assume a special case $a_s = 1$ \footnote{Note that the assumed value for $a_s$ chosen in such a way it does not violate the requirement of the classical notion of emergent cosmology. Since we are working with a natural unit system, $c = \hbar = 1$ and $G = 1/8\pi$, this gives the Planck length $\sqrt{1/8\pi}$. To avoid the quantum gravity era, the ES radius must be greater than the Planck length. Also, on a similar footing, the Planck energy density must be chosen such that $\rho_s < 64 \pi^2$}. This choice is taken in order to drastically simplify the existence regions. This gives
\begin{equation}
\lambda^2 = -\frac{8 \alpha  (\alpha +1)-4 \alpha  (2 \alpha +3) C^2+(6 \alpha +1) (\omega +1) \rho _s+2}{2 (2 \alpha +1)^2}
\label{lamsqk=1as=1}
\end{equation}
This form of eigenvalues squared is useful as it relates the energy density of the ESU with the model parameters $\alpha$ and $C$ along with the EoS $\omega$. From this relation, we may obtain the stability regions corresponding to the requirement of the model parameters. The sign of the eigenvalue squared ($\lambda^2$) determines the type of stability of the dynamical system. When $\lambda^2 < 0$, the critical point obtained for the ESU is a centre equilibrium point and has circular stability. In other words, any small perturbation from the critical point will lead to indefinite oscillations about the point rather than an exponential deviation from it. By simultaneously setting $\lambda^2 < 0$ required for stable solutions and $\rho_s > 0$, $\alpha > 0$ and $C>0$, one may obtain the constraints on $\omega$ as

\begin{equation}
\omega >\frac{-8 \alpha ^2-8 \alpha +8 \alpha ^2 C^2+12 \alpha  C^2-6 \alpha  \rho _s-\rho _s-2}{6 \alpha  \rho _s+\rho _s}
\label{omega_k=1}
\end{equation}
This constraint on $\omega$ is however not entirely informative about the explicit existence regions as it involves the unknown parameters in a coupled form. To reduce it further to a convincingly simpler form, we may need to eliminate $\rho_s$ and $\alpha$ (or $C$) from the expression. Also, note that the constraint on $\omega$ is not changed by the sign of $C$, which keeps the inequality \eqref{omega_k=1} unchanged irrespective of the sign of $C$. Now solving for $\rho_s$ from equation \eqref{freq3} we get
\begin{equation}
\rho_s = 3 \left(1 + \alpha - \alpha C^2\right)
\label{rhos_k1}
\end{equation}
Again setting $\rho_s > 0, \alpha >0$ the constraint on $C$ is obtained as
\begin{equation}
-\sqrt{\frac{\alpha +1}{\alpha }}<C<\sqrt{\frac{\alpha +1}{\alpha }}
\label{const_C}
\end{equation} 

This relation gives the possible ranges of $C$ given the values of $\alpha$. For $k = 1,$ with $\rho = \rho_s, a = a_s = 1, \dot{a} = \ddot{a} = 0$, the Raychaudhuri equation can be expressed in a simplified form for the ESU 
\begin{equation}
\frac{3}{2} (\omega +1) \left(\alpha -\alpha  C^2+1\right)+2 \alpha  C^2-(1+2\alpha) = 0,
\label{RC1}
\end{equation}
Solving for $\alpha$ gives 
\begin{equation}
\alpha = \frac{3 \omega +1}{\left(C^2-1\right) (3 \omega -1)},
\label{alp1}
\end{equation}
Keeping in mind the statement addressed earlier at the end of section \ref{intro}, we set $\alpha > 0$. Also, let us choose $C>0$ \footnote{Setting $C > 0$ or $C < 0$ would not have any effect on the stability regions as observed from \eqref{alp1}. We may choose $C>0$ conventionally.}, from which  we find the following constraints on the EoS parameter from the $\alpha > 0$ perspective for different ranges of $C$ 
\begin{equation}
\begin{aligned}
&\text{For } 0<C<1,  -\frac{1}{3} <\omega <\frac{1}{3}, \\
&\text{and for } C>1, \omega <-\frac{1}{3} \text{ or } \omega >\frac{1}{3}
\end{aligned}
\label{const_omeg_Ck=1}
\end{equation}
With the substitution of equations \eqref{rhos_k1} and \eqref{alp1} into the inequality \eqref{omega_k=1} we find the following existence regions for different ranges of $C$
\begin{widetext}
\begin{equation}
\begin{aligned}
&\left\{ (C, \omega) \mid 0 < C < \frac{1}{\sqrt{2}} \ \text{and} \ \left( \frac{-3 + C^2}{3 + 3C^2} < \omega < -\frac{5}{9} \ \text{or} \ \omega > \frac{-7 + C^2}{15 + 3C^2} \right) \right\} \\
&\cup \left\{ (C, \omega) \mid C = \frac{1}{\sqrt{2}} \ \text{and} \ \omega > -\frac{13}{33} \right\} \\
&\cup \left\{ (C, \omega) \mid \frac{1}{\sqrt{2}} < C \leq 1 \ \text{and} \ \left( -\frac{5}{9} < \omega < \frac{-3 + C^2}{3 + 3C^2}\ \text{or} \ \omega > \frac{-7 + C^2}{15 + 3C^2} \right) \right\} \\
&\cup \left\{ (C, \omega) \mid C > 1 \ \text{and} \ \left( -\frac{5}{9} < \omega < \frac{-7 + C^2}{15 + 3C^2}  \ \omega > \frac{-3 + C^2}{3 + 3C^2} \right) \right\}
\end{aligned}
\label{const_omeg_lamsq1}
\end{equation}
\end{widetext}
Note that these stability ranges are free from the dependence on $\rho_s$ and $\alpha$. Thus, the admitted parameter values for $\lambda^2 < 0$ lead to a centre equilibrium point and refer to a stable ESU. 

\subsection{Model 2: \(k = -1\)}
For the case of an open Universe, setting $k= -1$ in the Raychaudhuri equation \eqref{RCEqn}, and again considering $x_1 = a$ and $x_2 = \dot{x}_1 = \dot{a}$, the dynamical system for this case becomes
\begin{widetext}
\begin{equation}
\begin{aligned}
\dot{x}_1 &= x_2,\\
\dot{x}_2 &= -\frac{2 \alpha  x_1 \left(C^2+x_2^2-1\right)+4 \pi  G \rho  x_1^5 (\omega +1)+2 \alpha  x_2^2 x_1^2-2 \alpha  x_2^4-x_2 x_1^4+x_1^3}{2 \alpha  \left(x_2^2-1\right) x_1^2+x_1^4} \equiv \mathcal{X}(x_1,x_2).
\end{aligned}
\label{dynsysk=-1}
\end{equation}
\end{widetext}
The eigenvalue $\lambda$ for this system around the critical point $(x_1,x_2) = (a_s, 0)$ is 
\begin{equation}
\begin{aligned}
 \lambda^2 &= \frac{1}{2 \left(a_s^3-2 \alpha  a_s\right)^2} \left[ 4 \alpha  \left(3 C^2-2\right) a_s^2+a_s^4 \right. \\& \left. \left(6 \alpha  (\omega +1) \rho _s+2\right)-\left((\omega +1) a_s^6 \rho _s\right)-8 \alpha ^2 \left(C^2-1\right)\right]
 \end{aligned}
 \label{lamsqk=-1}
\end{equation}
Setting $a_s = 1$ (also setting $G = 1/8\pi$ simultaneously) the eigenvalue squared from equation \eqref{lamsqk=-1} becomes,
\begin{equation}
\lambda^2 = \frac{\begin{aligned} -8 \alpha ^2 \left(C^2-1\right)&+4 \alpha  \left(3 C^2-2\right)+(6 \alpha -1) \\&\hspace{2cm} (\omega +1) \rho _s+2 \end{aligned}}{2 (1-2 \alpha )^2}
\label{lamsqk=-1as=1}
\end{equation}
In this case, the energy density of the ESU becomes
\begin{equation}
\rho_s = 3(\alpha - 1 -\alpha C^2),
\label{rhos_k=-1}
\end{equation}
Setting $\rho_s >0$ in equation \eqref{rhos_k=-1} the constraint on $C$ becomes
\begin{equation}
-\sqrt{\frac{\alpha -1}{\alpha }}<C<\sqrt{\frac{\alpha -1}{\alpha }}, \quad \forall \, \alpha > 1.
\label{C_const_k=-1}
\end{equation}
Now, for stable solutions like the previous case the constraints on $\omega$ with $\lambda^2 < 0$ for $\rho_s > 0, \alpha > 1 $ and $C > 0$ are found as
\begin{equation}
\omega <\frac{-8 \alpha ^2+8 \alpha +8 \alpha ^2 C^2-12 \alpha  C^2-6 \alpha  \rho _s+\rho _s-2}{6 \alpha  \rho _s-\rho _s}
\label{omeg_const_k=-1}
\end{equation}
Substuting $\rho = \rho_s,$ from equation \eqref{rhos_k=-1} and $ \dot{a} = \ddot{a} = 0$ in equation \eqref{RCEqn} we get 
\begin{equation}
\frac{3}{2} (\omega +1) \left(\alpha -\alpha  C^2+1\right)+ 2 \alpha  C^2+(1-2\alpha) = 0,
\label{rc_eqn_k=-1}
\end{equation}
Solving \eqref{rc_eqn_k=-1} for $\alpha$ gives, 
\begin{equation}
\alpha = \frac{3 \omega +5}{\left(C^2-1\right) (3 \omega -1)}
\label{alp_k=-1}
\end{equation}
It is seen from equation \eqref{C_const_k=-1}, that real values of $C$ must require $\alpha > 1$. Imposing $\alpha>1$ and $C > 0$ the constraints on $C$ and $\omega$ are
\begin{equation}
\begin{aligned}
&\text{For } 0<C<1, \quad \frac{C^2+4}{3 C^2-6}<\omega <\frac{1}{3} \text{ or } \\
&\text{For } 1 <C<\sqrt{2}, \quad \left (\omega <\frac{C^2+4}{3 C^2-6}\,  \cup \, \omega >\frac{1}{3}\right) \text{ or } \\
&\text{ For } \left(C=\sqrt{2}, \quad \omega >\frac{1}{3}\right) \text{ or } \\
&\text{For } \left(C>\sqrt{2}, \quad \frac{1}{3}<\omega <\frac{C^2+4}{3 C^2-6}\right)
\end{aligned}
\label{const_omega_Ck=-1}
\end{equation}

Substituting equation \eqref{rhos_k=-1} and \eqref{alp_k=-1} into equation \eqref{omeg_const_k=-1} eliminates the dependence of $\omega$ on $\rho_s$ and $\alpha$ which gives the existence regions of stability (shown in appendix \ref{app1}).
As mentioned earlier, the sign of $C$ does not affect the analysis. This stability regions are true for both positive and negative values of $C$ provided the condition \eqref{C_const_k=-1} is met. 
\begin{figure*}[htb]
\centerline
\centerline{\includegraphics[scale=0.5]{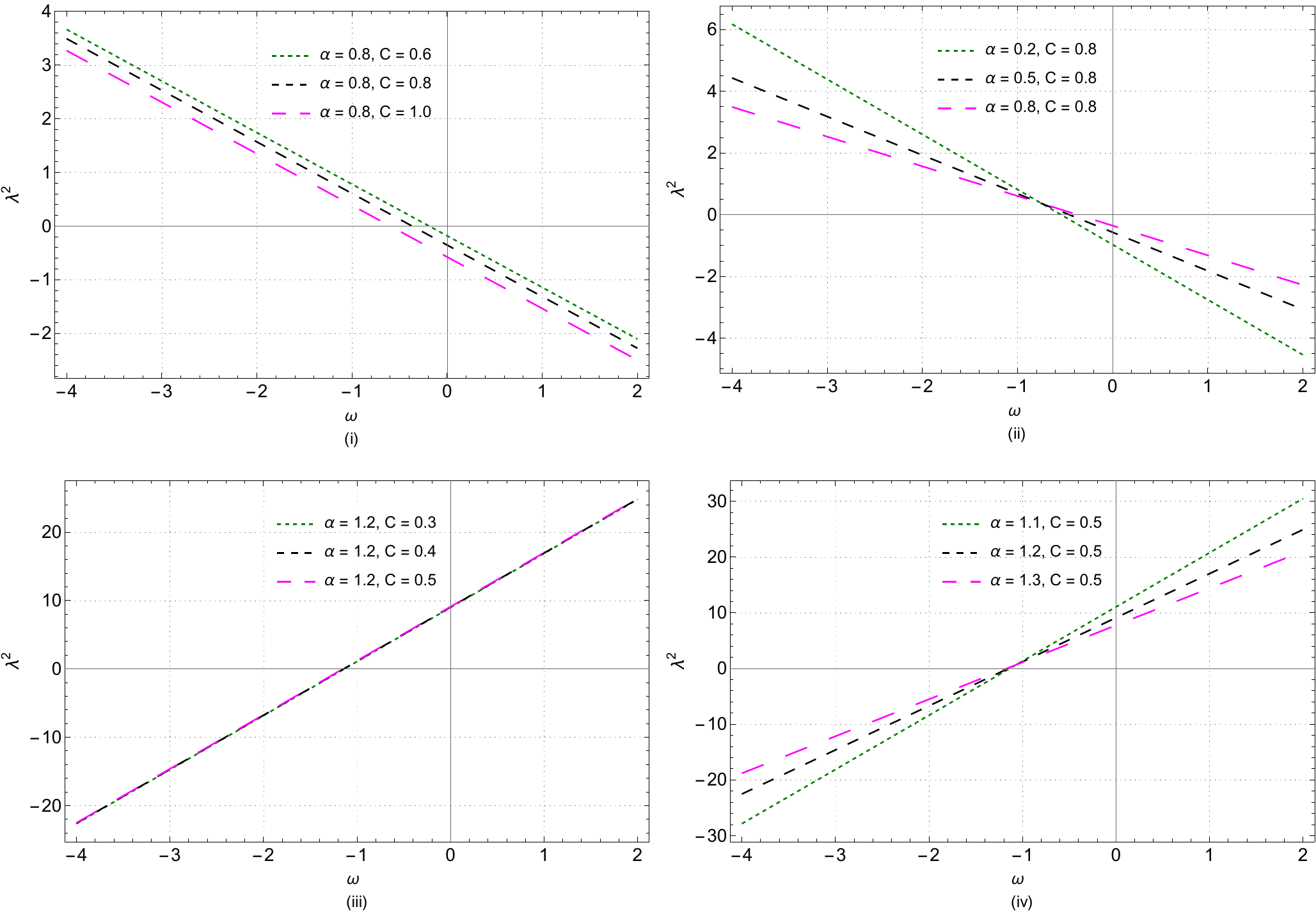}}
\caption{The evolution of $\lambda^2$ with $\omega$ is shown for different combinations of the parameter values for both $k=1$ (See (i) and (ii)) and $k = -1$ (See (iii) and (iv)) cases. }
\label{eigen_k=1}
\end{figure*}

Let us now address the graceful exit mechanism of the ESU to the standard cosmology for both scenarios of $k = 1$ and $k = -1$ by illustrating the evolution of the eigenvalue squared against the EoS parameter $\omega$. For $k = 1$, it is evident that $\lambda^2$ evolves linearly with decreasing $\omega$ from the negative to the positive region of $\lambda^2$. The phase transition occurs in the interval $w = [-1,0]$ for different sets of $(\alpha, C)$ parameter values, where $C$ is adjusted and $\alpha$ is kept fixed, as shown in Fig. \ref{eigen_k=1} (i).  Fig. \ref{eigen_k=1} (ii) shows a similar scenario where $\alpha$ varies but $C$ remains constant.
In contrast, when $k=-1$, it is found that as $\omega$ decreases, the variation of $\lambda^2$ is from the positive to the negative regions, regardless of different combinations of $(\alpha, C)$. Negative $\lambda^2$ indicates imaginary eigenvalues, which correspond to a stable centre where perturbations to the static solutions do not lead to a collapse or diverging of the solutions, resulting in a perpetual loop of oscillations around the ES critical point. However, $\lambda^2 > 0$ characterises a saddle-like point that indicates an unstable critical point. 
So, at this stage, one might anticipate finding a mechanism that overcomes these endless oscillations about the ESU critical point and leads to a subsequent unstable saddle point indicating the inflationary phase. The transition from negative to positive $\lambda^2$ may be viewed as a graceful exit from the stable ESU to the standard cosmological scenario, a fundamental requirement for emergent cosmology. 
\footnote{One may see that the decrease of $\omega$ from higher to lower numerical values may be interpreted as the overall growth of cosmic time. This interpretation is explained rigorously in \cite{Khodadi2018Jun}.} 
In recapitulation, $k=1$ displays promising behaviour while exhibiting a graceful exit from ESU to a standard inflationary scenario whereas $k=-1$ does not show the correct graceful exit mechanism. From the analysis, it is also clearly seen that $\omega$ plays the role of a bifurcation parameter, since depending on these values the qualitative behaviour of the ESU changes. 

The table of critical points for $k = 1$ and $k = -1$ are shown in TABLE. \ref{tab1} and TABLE. \ref{tab2} for different combinations of the model parameters. The critical points are obtained by setting $\rho_s = 1$.
\begin{table*}[tbh]
\centering
\begin{tabular}{||c|c|c|c|c|c||}
\hline
$k$                      & \textbf{$(\alpha, C)$}        & \textbf{$\omega$} & \textbf{Critical Points $(a_s, 0)$} & \textbf{$\lambda^2$} & \textbf{Stability} \\ \hline
\multirow{6}{*}{$k = 1$} & \multirow{3}{*}{$(0.8, 0.6)$} & $0.3$             & $(1.49, 0)$                         & $< 0$                & Stable             \\ \cline{3-6} 
 &                               & $-0.1$ & $(1.72, 0)$             & $< 0$  & Stable   \\ \cline{3-6} 
 &                               & $ -1$  & no real critical points & $ > 0$ & Unstable \\ \cline{2-6} 
 & \multirow{3}{*}{$(0.5, 0.8)$} & $0.3$  & $(1.53, 0)$             & $< 0$  & Stable   \\ \cline{3-6} 
 &                               & $-0.4$ & $(1.91, 0)$             & $< 0$  & Stable   \\ \cline{3-6} 
 &                               & $ -1$  & no real critical points & $ > 0$ & Unstable \\ \hline
\end{tabular}
\caption{Critical points for different combination of $(\alpha, C)$ and $\omega$ for $k = 1$}
\label{tab1}
\end{table*}

\begin{table*}[tbh]
\centering
\begin{tabular}{||c|c|c|c|c|c||}
\hline
$k$                       & \textbf{$(\alpha, C)$}        & \textbf{$\omega$} & \textbf{Critical Points $(a_s, 0)$} & \textbf{$\lambda^2$} & \textbf{Stability} \\ \hline
\multirow{6}{*}{$k = -1$} & \multirow{3}{*}{$(1.2, 0.3)$} & $0.3$             & $(0.824, 0)$                        & $ > 0$               & Unstable           \\ \cline{3-6} 
 &                               & $-1$    & $(1.47, 0)$             & $> 0$  & Unstable \\ \cline{3-6} 
 &                               & $ -1.3$ & no real critical points & $ < 0$ & Stable   \\ \cline{2-6} 
 & \multirow{3}{*}{$(1.1, 0.5)$} & $0.3$   & $(1, 0)$                & $ > 0$ & Unstable \\ \cline{3-6} 
 &                               & $-1$    & $(1.28, 0)$             & $ > 0$ & Unstable \\ \cline{3-6} 
 &                               & $ -1.3$ & $(1.91, 0)$             & $ < 0$ & Stable   \\ \hline
\end{tabular}
\caption{Critical points for different combination of $(\alpha, C)$ and $\omega$ for $k = -1$}
\label{tab2}
\end{table*}

\section{Stability under Homogeneous Linear Perturbation}
\label{sec4}
In this section, we aim to understand the stability of the ES Universe under linear homogeneous perturbations for $k = \pm 1$ Universes. Our motive is to find a possible influence of such perturbation on the stability of the ESU. A time-dependent perturbation is introduced into the scale factor and the energy density up to a linear order, given by
\begin{equation}
a(t) = a_0(1 + \delta a(t)), \quad \rho(t) = \rho_0 (1+ \delta \rho(t)),
\label{pert1}
\end{equation}
where $\delta a(t)$ and $\delta \rho(t)$ are infinitesimal linear perturbations introduced to the scale factor and energy density respectively.

For the ESU in the $k = 1$ case, setting $\dot{a} = \ddot{a} = 0$ and also $G = 1/8\pi$ in equation \eqref{freq3} and \eqref{RCEqn} we obtain
\begin{equation}
\rho_s d\rho = 3\left(\frac{4\alpha C^2}{a_s^4} - \frac{2}{a_s^2} - \frac{4\alpha}{a_s^4}\right) \delta a,
\label{rhodrho1}
\end{equation}
\begin{equation}
 \frac{1}{a_s^2} + \frac{\alpha}{a_s^4} - \frac{1}{3} \rho_s - \frac{\alpha C^2}{a_s^4} = 0
\label{perteq1}
\end{equation}
and for $k= -1$ case we get
\begin{equation}
\rho_s d\rho = 3\left(\frac{4\alpha C^2}{a_s^4} - \frac{2}{a_s^2} - \frac{4\alpha}{a_s^4}\right) \delta a,
\label{rhodrho2}
\end{equation}
\begin{equation}
- \frac{1}{a_s^2} + \frac{\alpha}{a_s^4} - \frac{1}{3} \rho_s - \frac{\alpha C^2}{a_s^4} = 0
\label{perteq2}
\end{equation}
Now, substituting the perturbed scale factor and the energy density in equation \eqref{freq3} and using the respective $\rho_s d\rho$ equations \eqref{rhodrho1} and \eqref{rhodrho2}, we finally obtain 
\begin{equation}
\delta \ddot{a} + \Omega_1 \delta a  = 0,
\label{perteq3}
\end{equation}
for $k = 1$ case and
\begin{equation}
\delta \ddot{a} + \Omega_2 \delta a  = 0,
\label{perteq4}
\end{equation}
for $k = -1$ case, where
\begin{equation}
\begin{aligned}
\Omega_1 &= \frac{3(1+\omega)}{2(a_s^2 + 2\alpha)}\left(\frac{4\alpha C^2 }{a_s^2} - \frac{4\alpha}{a_s^2} - 2\right),\\
\Omega_2 &= \frac{3(1+\omega)}{2(a_s^2 + 2\alpha)}\left(\frac{4\alpha C^2 }{a_s^2} - \frac{4\alpha}{a_s^2} + 2\right),
\end{aligned}
\label{Omeg1}
\end{equation}

For finite oscillating perturbation modes, which admit stable ES solutions, $\Omega_1, \Omega_2 > 0$. The stable solutions of the equations \eqref{perteq3} and \eqref{perteq4} are then
\begin{equation}
\delta a (t) = C_1 e^{i \Omega_1 t} + C_2 e^{-i\Omega_1 t},
\label{sol1}
\end{equation}
and
\begin{equation}
\delta a (t) = C_3 e^{i \Omega_2 t} + C_4 e^{-i\Omega_2 t},
\label{sol2}
\end{equation}
where $C_1$ and $C_2$ are integration constants. 
Therefore given the conditions $\Omega_1, \Omega_2 > 0$, the stability intervals are obtained to be
\begin{align*}
&\left\{ (C, \alpha, \omega) \mid 0 < C \leq 1 \ \text{and} \ \alpha > 0 \ \text{and} \ \omega < -1 \right\} \\
&\cup \left\{ (C, \alpha, \omega) \mid C > 1 \ \text{and} \ \left( 0 < \alpha < \frac{1}{-2 + 2C^2} \right. \right. \\& \left. \left. \hspace{2cm} \ \text{and} \ \omega < -1 \right. \right. \\
&\quad\quad \left. \left. \text{or} \ \alpha > \frac{1}{-2 + 2C^2} \ \text{and} \ \omega > -1 \right) \right\}
\end{align*}
for $k = 1$ whereas for $k = -1$ we obtain
\begin{align*}
&\left\{ (C, \alpha, \omega) \mid 0 < C < 1 \ \text{and} \ \left( 0 < \alpha < -\frac{1}{-2 + 2C^2} \ \text{and} \ \right. \right. \\& \left. \left. \hspace{2cm} \omega > -1 \right. \right. \\
&\quad\quad \left. \left. \text{or} \ \alpha > -\frac{1}{-2 + 2C^2} \ \text{and} \ \omega < -1 \right) \right\} \\
&\cup \left\{ (C, \alpha, \omega) \mid C \geq 1 \ \text{and} \ \alpha > 0 \ \text{and} \ \omega > -1 \right\}
\end{align*}
Like the previous analysis, we have also set $a_s = 1$ for simplicity. 

\begin{figure*}[tbh]
\centerline
\centerline{\includegraphics[scale=0.5]{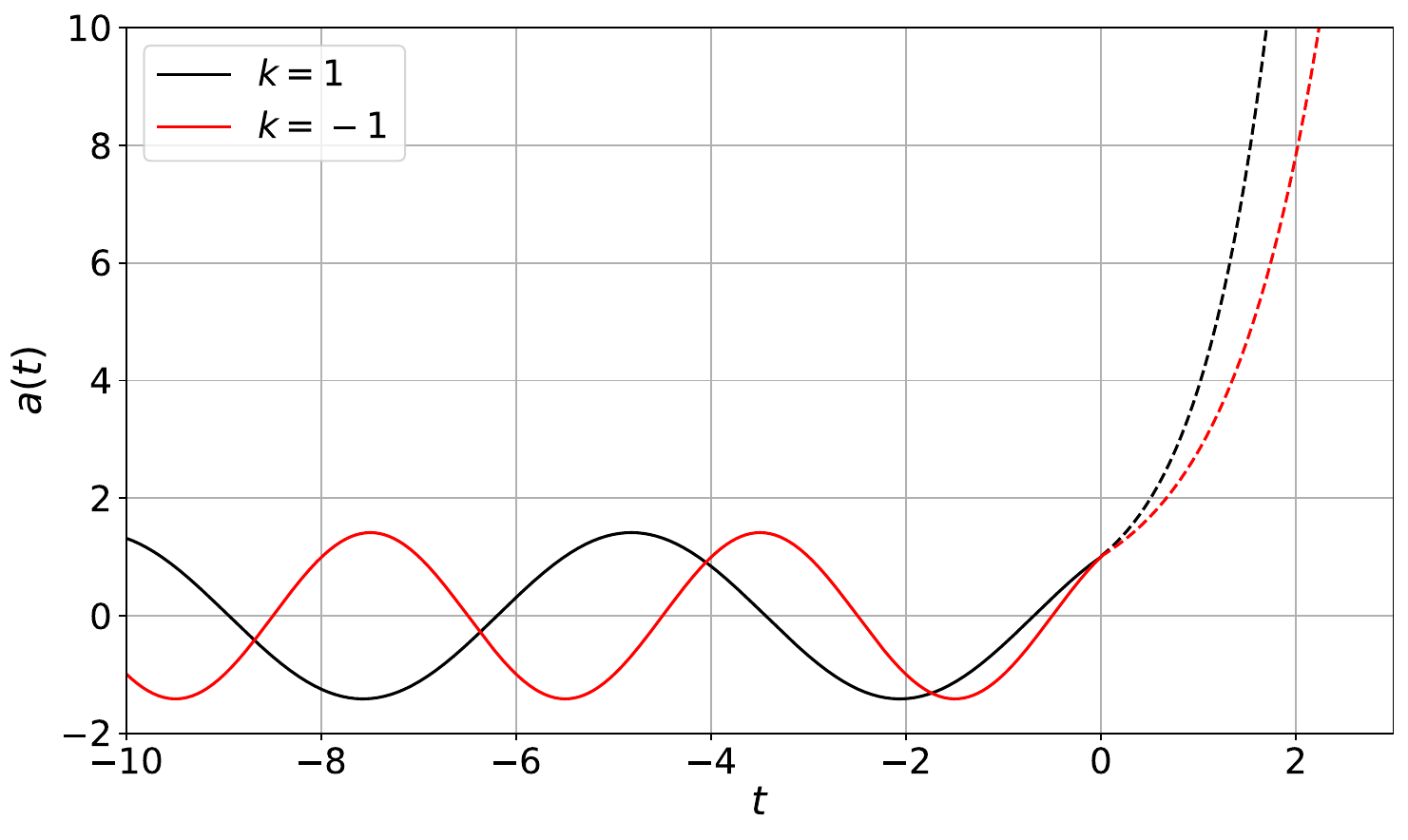}}
\caption{Graceful exit mechanism for $k=1$ and $k=-1$ under the influence of homogeneous perturbation}
\label{Grace_homo}
\end{figure*}

Figure \ref{Grace_homo} shows the graceful exit mechanism from a stable ES phase to the inflationary phase. In this analysis, we have set $t_0 = 0$ as the transition point. For $k=1$, setting $\omega = -0.3,\alpha = 0.5,C = 1.8$ gives the stable region of the ESU subject to homogeneous perturbation. However, a change in the value of $C$ from 1.8 to 0.5 breaks the infinite series of oscillations of the scale factor and leads to the exponential inflationary phase. But, for $\omega < -1$, the oscillations do not break and a graceful exit does not occur. Therefore, a phantom-like fluid does not allow a successful graceful exit from a closed Universe. For $k=-1$ case by setting $\omega = -0.3,\alpha = 1.5,C = 1.5$, we find that ESU is stable subject to homogeneous perturbation, but breaks when $\omega < -1$. This suggests that a successful graceful exit from ESU to the standard inflationary cosmology demands the requirement of a fluid of phantom nature ($\omega < -1$) for an open Universe. 

\section{Stability under Matter Perturbation}
\label{sec5}
The theory growth of matter density perturbation in the case of 4DEGB gravity was described by Haghani \cite{Haghani2020Dec} with a specific focus on the observed cosmological data. However, due to the generic nature of the perturbation equations, we may use them to study the evolution of the matter density contrast in the ESU as well. In the same context, Bohmer et al investigated the stability of ESU under perturbations in the Newtonian gauge in theories with scalar fluids \cite{Bohmer2015Dec}.

Following Haghani \cite{Haghani2020Dec}, we may write the perturbed conformal FLRW metric in the Newtonian gauge as 
\begin{equation}
ds^2 = a^2 (t) \left[ -(1+ 2\Phi) dt^2 + (1 - 2\Psi) dx_i ^2 \right],
\label{pertmet}
\end{equation}
where $\Phi$ and $\Psi$ denote the Newtonian gravitational potentials. 
Because of the fact that the conservation equation $\nabla_\mu T^{\mu \nu} =0$ takes the same form in both GR as well as in 4DEGB gravity, one may find the perturbed $00$-component of the metric field equation as \cite{Haghani2020Dec}
\begin{equation}
a^4 \rho \delta_m + 4 \mathcal{A} \left( k^2 \Psi + 3H^2 \Phi + 3H \dot{\Psi} \right) = 0,
\label{pert_field}
\end{equation}
where $\mathcal{A} = 2\alpha H^2 + a^2$\footnote{\cite{Haghani2020Dec} uses $\kappa$ in their field equations. We have set $\kappa = 1$ as we are using natural units.}

For both open and closed Universes, setting $k = \pm 1$  and the ES conditions $H = 0$,  $a = a_s$ and $\rho = \rho_s$ in \eqref{pert_field} leads to 
\begin{equation}
a_s^2 \rho_s \delta_m + 4 a_s^2 \Psi = 0,
\label{pert_esu}
\end{equation}
To solve for $\delta_m$ one needs to specify the Newtonian potential $\Psi$, which can be obtained by solving the Poisson's equation in the ESU given as 
\begin{equation}
\nabla^2 \Psi  = 4\pi G \rho_s,
\label{poisson}  
\end{equation}  
Assuming the ESU is spherically symmetric we have $\nabla^2 \equiv \frac{1}{a^2} \frac{\partial}{\partial a} (a^2  \frac{\partial}{\partial a} )$, using which we solve Eq. \eqref{poisson} that gives  
\begin{equation}
\Psi (a_s) = \frac{a_s^2 \rho_s}{12} + \frac{\mathbb{C}}{a_s}, 
\label{psi}
\end{equation}
where $\mathbb{C}$ is an integration constant. 
Thus, for an ESU  and keeping track of the previous analyses where $a = a_s =1$, Eq. \eqref{psi} takes the form
\begin{equation}
\Psi = \frac{\rho_s}{12} + \mathbb{C}.
\label{psi_esu}
\end{equation}
Setting Eq. \eqref{psi_esu} in Eq. \eqref{pert_esu} gives
\begin{equation}
\delta_m = - \frac{4}{\rho_s} \left( \frac{\rho_s}{12} + \mathbb{C} \right).
\label{delm}
\end{equation}
Now, subsituting the ES energy density for $k=1$ case from Eq. \eqref{rhos_k1} into Eq. \eqref{delm} we obtain 
\begin{equation}
\delta_m = \frac{\alpha -\alpha  C^2+5}{3 \alpha  \left(C^2-1\right)-3}
\label{delmk=1}
\end{equation}
and for $k = -1$ we use Eq. \eqref{rhos_k=-1} to obtain
\begin{equation}
\delta_m = \frac{\alpha -\alpha  C^2 + 3}{3 \alpha  \left(C^2-1\right)+3},
\label{delmk=-1}
\end{equation}
Eqs. \eqref{delmk=1} and \eqref{delmk=-1} clearly show that the ESU is neutrally stable against matter perturbation in the Newtonian gauge as these perturbations do not grow with time and stays finite at all times provided the conditions
\begin{equation}
\alpha \neq \frac{1}{C^2-1} \text{ for }  k = 1 \text{ and } \alpha \neq \frac{1}{1-C^2} \text{ for } k = -1,
\end{equation} are satisfied.

\section{Conclusion}
\label{conc}
In this paper, we have investigated emergent cosmology by studying the fundamental requirements of an emergent scenario namely \emph{stability of the ESU} and \emph{graceful exit mechanism}. The stability is analysed based on several techniques like dynamical systems, homogeneous scalar perturbations, density perturbations, vector perturbations and tensor perturbations. We assumed that the total matter content is a perfect fluid described by the constant equation of state $\omega$ with closed and open spatial geometry. From the dynamical system point of view, we have found that a closed Universe shows promising behaviour exhibiting stable ES solutions and providing a successful graceful exit into standard cosmology. However, a spatially open Universe, although stable ES solutions can be found does not show a successful graceful exit mechanism.

As far as linear homogeneous perturbation to the scale factor and energy density is concerned, the ESU is found to be stable under these perturbations. To realise standard cosmology, the stability of the ESU should break and exit into the inflationary era (graceful exit). It is observed that for $\omega < -1$, the oscillations do not break and a graceful exit does not occur. Therefore, a phantom-like fluid does not allow a graceful exit from a closed Universe. It suggests that for the realization of a successful graceful exit, $-1 < \omega < 0$. But in contrast, in the case of an open Universe, a successful graceful exit is only observed when $\omega < -1$. In other words, it requires a phantom equation of state for the successful realization of standard inflationary cosmology after a phase transition from the ESU in an open Universe. We find that tuning the value of the parameter $C$ initiates a phase transition from the stable ESU to the standard inflationary Universe, which advocates a successful graceful exit.
Finally, we also investigated the effect of matter perturbation in the Newtonian gauge on the stability of the ESU. We observe that under a direct condition between the model parameters $\alpha$ and $C$, the matter perturbations remain finite through out the ESU.
 
To conclude, the exact reason to why $C$ should be a dynamic degree of freedom that triggers the inflationary phase is not very clear in this paper and also currently beyond the scope of the paper. We plan to investigate this concern in future work.

\appendix
\section{Stability Region for $k=-1$} 
\label{app1}
\begin{widetext}
\begin{equation}
\begin{aligned}
&\left\{ (C, \omega) \mid 0 < C < 0.98995 \ \text{and} \ \left( \omega < -\frac{5}{3} \ \text{or} \ \frac{29 + C^2}{-21 + 3C^2} < \omega < \frac{19 - 3C^2}{3(-7 + C^2)} - \frac{2}{3} \sqrt{\frac{1 - 28C^2 + 4C^4}{(-7 + C^2)^2}} \right. \right. \\
&\qquad \left. \left. \text{or} \ \frac{19 - 3C^2}{3(-7 + C^2)} + \frac{2}{3} \sqrt{\frac{1 - 28C^2 + 4C^4}{(-7 + C^2)^2}} < \omega < -\frac{2}{3} \right) \right\} \\
&\cup \left\{ (C, \omega) \mid C = 0.98995 \ \text{and} \ \left( \omega < -\frac{5}{3} \ \text{or} \ \frac{29 + C^2}{-21 + 3C^2} < \omega < \frac{19 - 3C^2}{3(-7 + C^2)} - \frac{2}{3} \sqrt{\frac{1 - 28C^2 + 4C^4}{(-7 + C^2)^2}} \right. \right. \\
&\qquad \left. \left. \text{or} \ \frac{19 - 3C^2}{3(-7 + C^2)} - \frac{2}{3} \sqrt{\frac{1 - 28C^2 + 4C^4}{(-7 + C^2)^2}} < \omega < -\frac{2}{3} \right) \right\} \\
&\cup \left\{ (C, \omega) \mid 0.98995 < C \leq 1 \ \text{and} \ \left( \omega < -\frac{5}{3} \ \text{or} \ \frac{29 + C^2}{-21 + 3C^2} < \omega < -\frac{2}{3} \right) \right\} \\
&\cup \left\{ (C, \omega) \mid 1 < C \leq 1.05573 \ \text{and} \ \left( \omega < \frac{29 + C^2}{-21 + 3C^2} \ \text{or} \ -\frac{5}{3} < \omega < -\frac{2}{3} \right) \right\} \\
&\cup \left\{ (C, \omega) \mid 1.05573 < C < \sqrt{7} \ \text{and} \ \left( \omega < \frac{29 + C^2}{-21 + 3C^2} \ \text{or} \ -\frac{5}{3} < \omega < -\frac{2}{3} \right. \right. \\
&\qquad \left. \left. \text{or} \ \frac{19 - 3C^2}{3(-7 + C^2)} - \frac{2}{3} \sqrt{\frac{1 - 28C^2 + 4C^4}{(-7 + C^2)^2}} < \omega < \frac{19 - 3C^2}{3(-7 + C^2)} + \frac{2}{3} \sqrt{\frac{1 - 28C^2 + 4C^4}{(-7 + C^2)^2}} \right) \right\} \\
&\cup \left\{ (C, \omega) \mid C = \sqrt{7} \ \text{and} \ \left( -\frac{5}{3} < \omega < -\frac{2}{3} \ \text{or} \ \omega > \frac{25}{3} \right) \right\} \\
&\cup \left\{ (C, \omega) \mid C > \sqrt{7} \ \text{and} \ \left( \omega < \frac{19 - 3C^2}{3(-7 + C^2)} - \frac{2}{3} \sqrt{\frac{1 - 28C^2 + 4C^4}{(-7 + C^2)^2}} \ \text{or} \ -\frac{5}{3} < \omega < -\frac{2}{3} \right. \right. \\
&\qquad \left. \left. \text{or} \ \frac{19 - 3C^2}{3(-7 + C^2)} + \frac{2}{3} \sqrt{\frac{1 - 28C^2 + 4C^4}{(-7 + C^2)^2}} < \omega < \frac{29 + C^2}{-21 + 3C^2} \right) \right\}
\end{aligned}
\label{const_omeg_lamsq-1}
\end{equation}
\end{widetext}

\bibliography{bibliography.bib}
\end{document}